\documentclass[aps,groupedaddress,floatfix]{revtex4}
\usepackage{graphicx}
\usepackage{float}
\usepackage[dvips]{color}
\usepackage[latin1]{inputenc}
\def\bc{\begin{center}}
\def\ec{\end{center}}

\def\beq{\begin{equation}}
\def\eeq{\end{equation}}

\def\d{\downarrow}
\def\u{\uparrow}

\def\bj{{\bf j}}

\begin{document}

%Title of paper
\title{Inelastic scattering of atoms in a double well}

\author{E. S. Annibale$^{1,2}$}
\email{annibale@if.usp.br}
\author{O. Fialko$^{2,3}$}
\email{Oleksandr.Fialko@physik.uni-augsburg.de} 
\author{K. Ziegler$^{2}$}
\email{klaus.ziegler@physik.uni-augsburg.de}
\affiliation{
$^1$Instituto de F\'isica, Universidade de S\~ao Paulo,  05508-090, 
S\~ao Paulo, Brazil\\
$^2$Institut f\"{u}r Physik, Universit\"{a}t Augsburg, D-86135, 
Augsburg, Germany\\
$^3$Centre for Theoretical Chemistry and Physics, Massey University (Albany 
Campus),
Private Bag 102904, North Shore MSC, Auckland 0745, New Zealand
}

\date{\today}

\begin{abstract}
We study a mixture of two light spin-1/2 fermionic atoms and two heavy atoms %in a Mott state 
in a double well potential. 
%The light atoms can tunnel and scatter inelastically with the heavy atoms. 
Inelastic scattering processes between
both atomic species excite the heavy atoms and renormalize the tunneling rate and the
interaction of the light atoms (polaron effect). %as well as the interaction of the light atoms.
The effective interaction of the light atoms changes its sign and becomes attractive for 
strong inelastic scattering.
This is accompanied by a crossing of the energy levels from singly occupied sites at weak 
inelastic scattering to a doubly occupied and an empty site 
for stronger inelastic scattering.
We are able to identify the polaron effect and the level crossing in the quantum dynamics.
\end{abstract}

\maketitle

%\pacs{03.75.Ss, 03.75.Mn, 34.50.-s, 71.10.Fd}

%\maketitle
%%%%%%%%%%%%%%%%%%%%%%%%%%%%%%%%%%%%%%%%%%%%%%%%%%%%%%%%%%%%%%%%%%%%%%%%%%%%%%%%

\section{Introduction}

Ultracold bosonic and fermionic atomic gases in optical 
lattices can be used as a toolkit for the investigation of fundamental 
condensed matter physics models \cite{lewen07}. 
Recent experimental work opened a field to study quantum states in optical 
lattices, such as superfluid and Mott states \cite{greiner02,bloch05}, 
where the interparticle interaction can be controlled  by a magnetic field via a
Feshbach resonance \cite{feshb}.  Spin-dependent effects 
\cite{ketterle03,mandel03,schmaljohann04,kohl05,partridge06}, frustrated spin 
systems \cite{lewenstein06}, the formation of dimers from fermionic atoms 
\cite{hulet03,uys05}, and mixtures of two atomic species 
\cite{stan04,esslinger06,sengstock06} provide opportunities for creating and 
studying even more complex quantum states.

The optical lattices are robust and free of phonons. On the other hand, the 
electron-phonon interaction in a solid leads to a rich physics. 
It is important for superconductivity, the Peierls instability, polaron effects
and many other phenomena. With more progress in the atomic and laser
physics, the coupling of ultracold atoms in an optical lattice to bosonic degrees 
of freedom may be achieved and thus can mimic the dynamics of electrons in the 
presence of phonons.
Recently, ultracold atoms confined to an optical resonator were proposed to study 
the effect of coupling between the atoms and the photons field, which leads to an 
effective Hubbard Hamiltonian with long-range interaction \cite{Ritsch05} and to 
an interesting phase diagram \cite{lewen06}. Bose-Fermi mixtures can also 
provide an insight into the role of bosons in the dynamics of fermions, 
where the condensed bosons lead to fermionic pairing \cite{wang06} and charge 
density waves of fermions \cite{hofst08}. 
An interesting example of the latter are dimer states. 
They have been discussed in solid-state systems \cite{rokhsar88}, in the Holstein-Hubbard
model \cite{ziegler08} and recently 
also for an ultracold Bose gas with ring exchange \cite{xu06}.

More recently, ultracold gases were employed to study the dynamics 
of quantum states, including the ``collapse and revival'' behavior \cite{greiner02b}.
Here it is important to distinguish between small systems with a few atoms and
many-body systems with a large number of atoms
% have attracted interest by the condensed matter and quantum computational communities 
\cite{lewen07}.  
For instance, it was observed experimentally that in a small system with two spin-1/2 atoms 
the spin dynamics and the particle dynamics are completely separated, similar to the
spin-charge separation in one-dimensional systems \cite{bloch07,bloch08}.
Another example for a restricted dynamics in small atomic systems are
entangled squeezed states in a Bose-Einstein condensate, whose atoms are
distributed over a small number of lattice sites \cite{ober08}. Both observations
indicate that the dynamics of small atomic systems can be restricted to a subspace of 
the entire Hilbert space available for the model Hamiltonian. This can also mean that the
system never reaches the groundstate of the Hamiltonian if it was prepared in an
excited state, reflecting the fact that the initial state has no overlap with the
groundstate. For two spin-1/2 atoms in a double well, described by a Hubbard model, 
this is a direct consequence of the fact that the eigenstates do not mix pairs
of singly occupied sites with pairs of empty and doubly occupied sites \cite{ziegler09}.
On the other hand, mixing of these states in a macroscopic system, enforced by
inelastic scattering with other atoms, can lead to
a first-order quantum phase transition from singly occupied sites to doubly
occupied sites. This was observed in a mixture of light and heavy atoms, where
the latter are in a Mott state \cite{ziegler08}. This case can be described
by a Bose-Fermi model that is
known in solid-state physics as the Holstein-Hubbard (HH) model \cite{alexandrov95}.
Adjusting physical parameters, such as the optical-lattice parameters 
(frequency and amplitude of the Laser field) and the fermion-fermion
interaction through a Feshbach resonance, 
enables us to prepare such a system not only in the ground state but also 
in its excited states and to study its dynamics. Although in a small system 
there is no phase transition for the ground state, dynamical properties
of excited states can change qualitatively due to inelastic scattering
described by the HH model. Among other effects, there is renormalization
of the tunneling rate, known as the polaron effect, which was recently also
discussed for an ultracold Bose gas \cite{tempere09}.

Our interest is to study the effect of inelastic scattering of two spin-1/2 fermionic
atoms in a double-well potential with repulsive Hubbard interaction and an additional
scattering with a heavy atom in each potential well. Each of the heavy atoms
have a harmonic oscillator spectrum and can exchange energy with the tunneling spin-1/2 atoms.

\begin{figure}[h!!]
\begin{center}
\includegraphics[width=7cm]{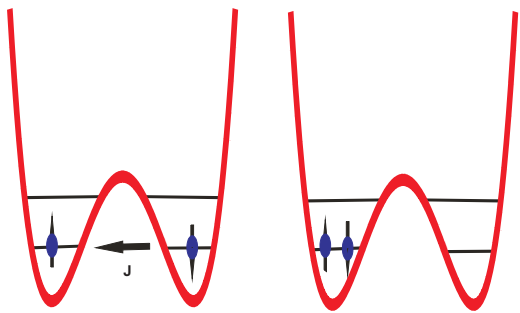}
\caption{\label{draft} Two LFA in a double well potential with one HA in each well. 
The HA are harmonic oscillators and are represented by horizontal lines 
(energy levels). 
The LFA can tunnel from one well to the other one. 
The wells can be singly occupied ({\bf left panel}) or one well can be doubly occupied 
and the other one empty ({\bf right panel}).}
\end{center}
\end{figure}

In this paper, we study a double well potential  
filled with heavy (bosonic or fermionic) atoms (HA)
(e.g. $^{87}$Rb or $^{40}$K), one at each  well (see Fig. \ref{draft}). 
A double-well potential can be realized by superimposing two periodic potentials
with different periodicity and the dynamics of one or two atoms can be easily
studied.
The tunneling of these HA is neglected since the potential
barrier between the wells is sufficiently high. Excitations are only due to
collisions with other atoms. For this purpose two light fermionic atoms  (LFA) 
(e.g. $^6$Li), prepared in two hyperfine 
states denoted as $|\!\uparrow\rangle$, $|\downarrow\rangle$, 
are added to the system. These atoms can tunnel because of their low mass
and scatter by the HA. It is assumed that the HA experience a
harmonic potential of a well, at least at low energies. 
Then their excitations are harmonic-oscillator states. During the
scattering process the HA can also transfer energy to the LFA.
Moreover, the light fermions experience local (on-site) repulsion.

The paper is organized as follows. The Holstein-Hubbard model is introduced 
and discussed in Sect. \ref{sec2}. 
In Sect. \ref{sect.restricted} we introduce a restricted model with at most
one phonon excitation per well.
Then in Sect. \ref{sect3} the effective Hamiltonian of the unrestricted 
Holstein-Hubbard model is defined, its spectral properties are studied and compared with those of
the restricted model of Sect. \ref{sect.restricted}. Based on this effective Hamiltonian we
study the dynamics of the quantum states in a double well, the spectral density and the
spin imbalance in Sect. \ref{sect4}.

%%%%%%%%%%%%%%%%%%%%%%%%%%%%%%%%%%%%%%%%%%%%%%%%%%
\section{Holstein-Hubbard Model}
\label{sec2}

The atomic mixture of LFA and HA can be well described by the 
Holstein-Hubbard model \cite{alexandrov95}:
%\linebreak
\begin{equation}\label{h_t}
H = -J\sum_{<\bj,\bj'>}\sum_{\sigma=\u,\d}
c_{\bj\sigma}^\dagger c_{\bj'\sigma}^{\phantom{a b}}+h.c.
+\sum_{\bj}\Big[\omega_0 b_\bj^\dagger b_\bj
+g(b_\bj^\dagger+b_\bj)(n_{\bj\u}+n_{\bj\d})
+Un_{\bj\u}n_{\bj\d}\Big]  \hspace{0.2cm} .
\end{equation}
The first term describes the tunneling of LFA with spin $\sigma$ ($=\u,\d$)
between nearest-neighbor wells.
These are defined by fermionic creation and annihilation 
operators $c_{\bj\sigma}^\dagger$ and  $c_{\bj\sigma}$, respectively.
The HA form a Mott state and are presented as harmonic oscillators 
at each well with eigenfrequency $\omega_0$,
assuming that a HA in one well is excited independently of the HA 
in the other well.
Thus they can be considered as local phonons and are described by the 
bosonic creation and annihilation operators 
$b_\bj^\dagger$ and $b_\bj$. 
The phonons couple to the light atoms with
strength $g\sim \langle e | \hat{V} | f \rangle$, where $\hat{V}$ 
is the interaction between LFA and HA, $|f \rangle$ denotes the 
ground state of a HA, while $| e \rangle$ denotes its first excited state.
The fourth term describes the interaction between two LFA at the same well,
where $U$ %=g_0\int w^4({\bf{x}}){\rm d} {\bf{x}}$ 
is a local repulsive interaction between the LFA

This lattice model describes the quantum phase transition in a half-filled system 
from singly occupied lattice wells (N\'eel state) to a mixture of doubly occupied 
and empty wells (dimer state). Now we restrict the lattice model to the two sites
of the double well potential, choosing the coordinates $j=1,2$ for the wells.
Ignoring the tunneling of the LFA and applying a unitary transformation to the 
remaining part of the Hamiltonian we can decouple fermionic and bosonic degrees 
of freedom and get the transformed local Hamiltonian  \cite{ziegler08}
\beq
H_\gamma = \sum_{\bj = 1,2}
\left[ \omega_0 b^\dagger_\bj b_\bj
+\gamma n_{\bj\u} n_{\bj\d}
-{g^2\over\omega_0}(n_{\bj\u}+n_{\bj\d}) \right]  \hspace{0.2cm}
\label{newham0}
\eeq
with $\gamma=U-2g^2/\omega_0$ is the effective Hubbard coupling. 
For a system with two fermions the ground state energy of $H_\gamma$ 
is given by
\beq
E_0=\epsilon - \gamma j \hspace{0.2cm},
\label{ground0}
\eeq
with $\epsilon = U - 4g^2/\omega_0$ and
$2 j$ is the number of singly occupied wells.
The coupling $\gamma$ controls two different regimes: for $\gamma > 0$, 
the ground state has two singly occupied wells ($j = 1$) 
and energy $-2g^2/\omega_0$, while for $\gamma < 0$, there are one doubly 
occupied well and one empty well ($j = 0$) 
and the energy is $\epsilon $. Thus at $\gamma = 0$ there is a transition, 
where the system changes from
two singly occupied wells to a doubly occupied well and an empty well.
Moreover, both states are degenerate even for $\gamma\ne0$
because the {\it local} Hamiltonian $H_\gamma$ does not determine how the spins, 
and the empty and doubly occupied wells are distributed in the 
double-well potential.
Tunneling in the Hamiltonian $H$ lifts these degeneracies \cite{alexandrov95}.
\begin{figure}[t]
\begin{center}
\includegraphics[width=7cm]{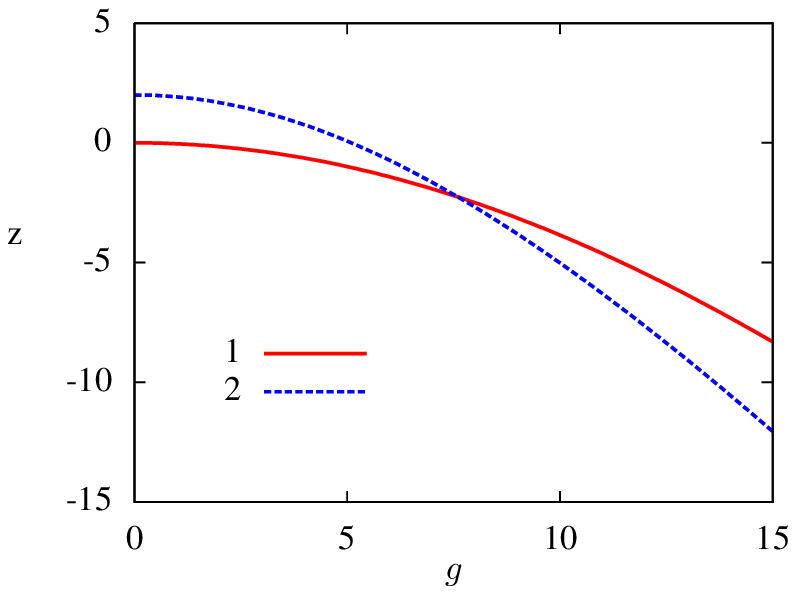}
\includegraphics[width=7cm]{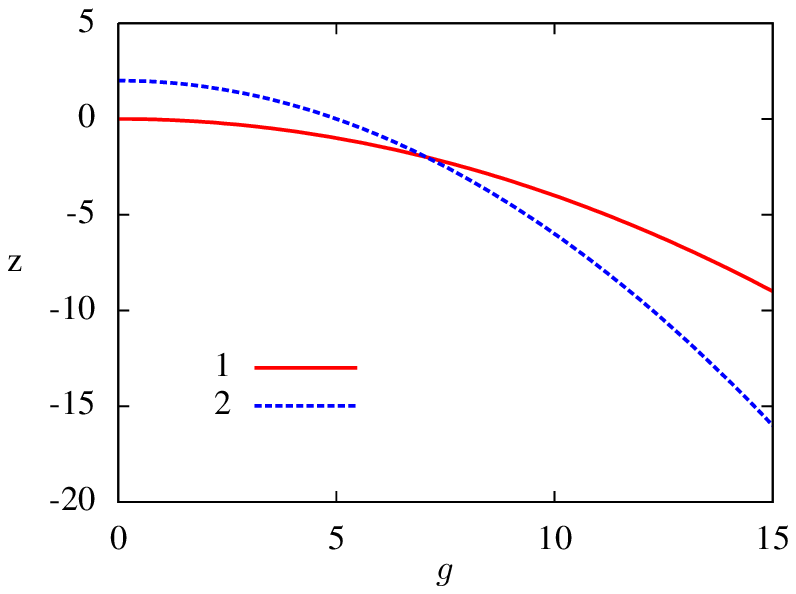}
%{figJ0.eps}
\caption{\label{fig2b}
Model without tunneling ($J=0$):
Four degenerate eigenvalues of the two-sites problem with at most 
one phonon excitation ({\bf left panel})
and the four lowest poles of the resolvent of Eq. (\ref{resolv}) 
with the effective Hamiltonian $H_{\rm eff}$ in Eq. (\ref{Heff})
({\bf right panel}) for $U = 2$ and $\omega_0 = 50$. Curve 1 (curve 2) represent
levels with two singly (one empty and one doubly) occupied wells.
{\bf Left panel}: The two curves cross at $g \approx 7.6$.
The energy of the ground state has a cusp, 
but it does not coincide with the exact ground state for $J=0$ 
given in Eq. (\ref{ground0}).
{\bf Right panel}: The two curves cross at the smaller value $g = 5 \sqrt{2}\approx 7.07$.}
\end{center}
\end{figure}
\begin{figure}[t]
\begin{center}
\includegraphics[width=7cm]{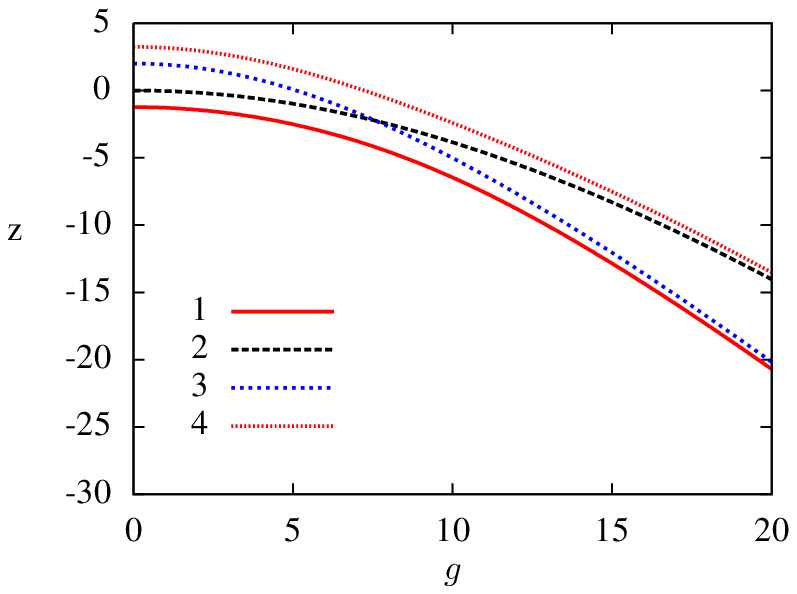}
\includegraphics[width=7cm]{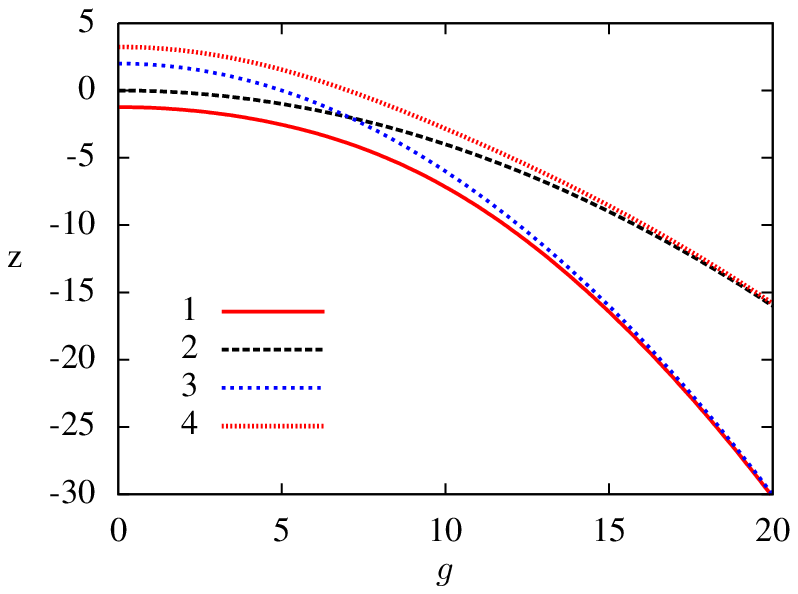}
%{poles_proj22.eps}
\caption{\label{fig3b}
Model with tunneling ($J=1$):
Four eigenvalues of the two-sites problem with at most one phonon excitation ({\bf left panel})
%%%%
and the four lowest poles of the resolvent of Eq. (\ref{resolv}) with the effective 
Hamiltonian $H_{\rm eff}$ of Eq. (\ref{Heff}) ({\bf right panel})
%%%%
for $U = 2$ and $\omega_0 = 50$. 
{\bf Left panel}: The curve 2 and the curve 3 cross at $g \approx 7.6$. For very large values 
of coupling $g$ the distances between curve 1 and curve 3, as well as between 
curve 2 and curve 4, remain nonzero.
{\bf Right panel}: The curve 2 and the curve 3 cross at $g \approx 7.1$. 
For very large values of coupling $g$ the distances between curve 1 and curve 3, 
as well as between curve 2 and curve 4, vanish as $\sim \tau^2/|\gamma|$. 
This is due to the polaron effect, which accounts for rescaling
of the single fermion tunneling $J\rightarrow \tau = J e^{-g^2/\omega_0^2}$.}
%This is due to the polaronic effect, which accounts for rescaling
%of the single fermion tunneling $J\rightarrow \tau = J e^{-g^2/\omega_0^2}$.}
\end{center}
\end{figure}

\section{Double well with at most one phonon excitation per well}
\label{sect.restricted}

Even for a double well with two fermions the Hilbert space of the Hamiltonian in Eq. (\ref{h_t}) 
has infinite dimensions due to the phonon excitations. For nonzero tunneling rate $J$ 
this becomes a difficult problem. However, to study qualititatively the effect of
inelastic scattering we can restrict the phonon excitations. The simplest case is a
model in which each of the localized atoms is a two-level system, similar to the 
Jaynes-Cummings model in quantum optics \cite{jaynes63}. 
For $J=0$ the four lowest eigenvalues are degenerated in pairs, as shown in the left panel of Fig. \ref{fig2b}.
The ground state has a cusp, but the latter does not coincide with the exact ground state 
for $J=0$ given in Eq. (\ref{ground0}). A nonzero tunneling rate $J$ lifts the degeneracy
and all four eigenvalues are distinct now (cf. left panel of Fig. \ref{fig3b}). 
We notice that for large values of $g$ the curves do not merge 
(i.e., the distances between the first and the third, and  the second 
and fourth curves, respectively, remain nonzero).

\section{Effective Hamiltonian for many phonon excitations}
\label{sect3}

Now we consider the full Holstein-Hubbard Hamiltonian in Eq. (\ref{h_t})
and treat its spectral properties in an approximative manner. The main
idea is to study the evolution of the quantum system 
$|\Psi_t\rangle=e^{-itH}|\Psi_0\rangle$, beginning with the initial state
$|\Psi_0\rangle$. The evolution is a walk through the entire Hilbert space
that is accessible for the Hamiltonian.
The Recursive Projective method (RPM) organizes this walk by projecting
iteratively on a sequence of subspaces. The main advantage of this method
is that the walk visits each subspace only once 
\cite{ziegler03,ziegler06,ziegler09}. The approximation method within
the RPM consists of ignoring some part of the Hilbert space that contributes
with a low probability to the walk and leads to an effective Hamiltonian.
Details of the application of the RPM
to the Holstein-Hubbard Hamiltonian can be found in Ref. \cite{ziegler08}.   
In the following we start from the effective Hamiltonian that was derived 
in Ref. \cite{ziegler08}, 
to study the dynamics of the LFA in the double-well potential. 
The advantage of this method is that it enables us to study the effect of finite 
tunneling of the LFA as well as arbitrary number of phonon excitations.

In order to derive an effective Hamiltonian, we project the full Hilbert space of
the Hamiltonian in Eq. (\ref{h_t}) onto the Hilbert space spanned by the four Fock states 
$|\!\uparrow,\downarrow\rangle$ , $|\downarrow,\uparrow\rangle$ ,
$|0,\uparrow\downarrow\rangle$
and $|\!\uparrow\downarrow, 0\rangle$ with a projector $P_0$, 
such that the resolvent of the projected Hamiltonian is given by
\begin{equation}\label{resolv}
G_0(z) = P_0 (z-H)^{-1}P_0 = \bigl(z - H_{\rm eff}(z)\bigl)^{-1}_0 \hspace{0.2cm}.
\end{equation}

The effective Hamiltonian for the double well can be evaluated recursively by the RPM. 
Under the assumption that the tunneling rate of LFA $J$ is small
compared to the other parameters of the system 
(e.g. $J \ll \omega_0$, $J \ll U$ and $J \ll g$) the recursion relation can be truncated, 
which gives \cite{ziegler08}
\begin{equation}
H_{\rm eff}(z)
\approx\sum_{\sigma,\sigma'=\uparrow,\downarrow}%\sum_{<\bj,\bj'>}
\left[ -\tau 
c^\dagger_{1\sigma}c_{2\sigma}
+ K_1(z)c_{1 \sigma}^\dagger c_{2\sigma}
        c_{2\sigma'}^\dagger c_{1\sigma'}\right. 
\left. 
+ K_2(z)c_{1 \sigma}^\dagger c_{2\sigma}
c_{1 \sigma'}^\dagger c_{2\sigma'}+h.c. +\frac{\gamma}{2} 
(n_{1\sigma}n_{1\sigma'}+n_{2\sigma}n_{2\sigma'})
\right],
\label{Heff}
\end{equation}
where the indices 1 and 2 represent the left and right sites 
of the double well, respectively.
% R/L denotes the right and left wells, respectively.
This Hamiltonian describes three different tunneling processes, namely
the tunneling of single fermions with rate $\tau$ (first term), 
the exchange of spins with rate $K_1(z)$ (second term), the tunneling of 
fermionic pairs with rate  $K_2(z)$ (third term) and the on-site interaction 
between fermions with strength $\gamma$ (fourth term). 
The tunneling rate $J$ of single fermions in Eq. (\ref{h_t}) is now renormalized as
\beq
\tau=e^{-g^2/\omega_0^2}J \ ,
\label{ren1}
\eeq
which is the well-known polaron effect \cite{alexandrov95}. The spin-exchange parameter $K_1$ and 
the pair tunneling parameter $K_2$ are given by the expressions \cite{ziegler08}
\begin{equation}\label{pra24a}
K_1(z)~=~2 \tau^2 \sum_{m = 1}^\infty \frac{1}{m!}
        \frac{\bigl( 2 g^2/\omega_0^2 \bigr)^m}
             {z \, - \, 2 \epsilon \, + \, 2 \gamma \, - \, \omega_0 m}
\end{equation}
and
\begin{equation}\label{pra24b}
K_2(z)~=~2 \tau^2\sum_{m = 1}^\infty \frac{1}{m!}
        \frac{\bigl( -2 g^2/\omega_0^2 \bigr)^m}
             {z \, - \, 2 \epsilon \,- \, \omega_0 m}.
\end{equation}
In order to avoid the singularities of the coefficients $K_1$ and $K_2$, we assume that 
$\omega_0 \gg U, g$, % and $\omega_0 \gg g$.
which is valid for a deep and tight double-well potential.

The energy levels of the system is given by the poles of Eq. (\ref{resolv}).
Thus the variable $z$ is fixed by solving the equation
\beq
\det [z - H_{\rm eff}(z)]=0.
\label{app2.poles}
\eeq
To solve this equation we first diagonalize the $4\times4$ effective Hamiltonian $H_{\rm eff}$  for a fixed 
parameter $z$ and find its eigenvalues $\lambda_j(z)$ ($j=1,...,4$). 
Then we solve $z = \lambda_j(z)$ for each of the four eigenvalues $\lambda_j(z)$ 
to determine the poles of the resolvent in Eq. (\ref{resolv}).
An eigenstate $|E\rangle$ (with $H_{\rm eff}|E\rangle=E|E\rangle$)
can be written as a linear combination of the four Fock states as
\begin{equation}
|E\rangle=
a_1(z)|\!\uparrow,\downarrow\rangle + a_2(z)|\!\downarrow,\uparrow\rangle +
a_3(z)|0,\uparrow\downarrow\rangle + a_4(z)|\!\uparrow\downarrow, 0\rangle  
\hspace{0.2cm} ,
\label{expansion}
\end{equation} 
where the coefficients $a_j(z)$ run over all possible poles $z$.
In this Fock-state basis the Hamiltonian in Eq. (\ref{Heff}) reads  
%$P_0$-projected Hilbert space the Hamiltonian in Eq. (\ref{Heff}) reads
\begin{equation}\label{Hmatrix}
{\hat H}_{\rm eff}(z) = \left(
\begin{array}{cccc}
K_1 + \epsilon - \gamma& -K_1 & -\tau &-\tau\\
-K_1 & K_1 +\epsilon - \gamma&\tau &\tau\\
-\tau &\tau&K_1 +\epsilon& K_2\\
-\tau &\tau&K_2& K_1 + \epsilon 
\end{array}
\right)\hspace{0.2cm},
\end{equation}
whose eigenvalues 
%of matrix (\ref{Hmatrix})
and the coefficients of the corresponding (non-normalized) eigenvectors 
in the form of Eq. (\ref{expansion}) are for doubly occupied lattice sites
\begin{equation}\label{lamb1}
\lambda_1(z) = K_1(z) - K_2(z) + \epsilon :\ \ \ 
%\end{equation}
%\begin{equation}\label{evec1}
a_1 = a_2 = 0 \, ; \hspace{0.5cm}
a_3 = -1 \, ; \hspace{0.5cm} a_4 = 1
\end{equation}
and for singly occupied lattice sites
\begin{equation}\label{lamb2}
\lambda_2(z) =-\gamma + \epsilon: \ \ \
%\end{equation}
%\begin{equation}\label{evec2}
a_1 = a_2 = 1 \, ; \hspace{0.5cm}
a_3 = a_4 = 0
\ .
\end{equation}
There are also states with a mixture of singly and doubly occupied sites:
Using $\Xi = \sqrt{16 \tau^2 + (-K_1(z) + K_2(z) + \gamma)^2}$ we have
\begin{equation}\label{lamb3}
\lambda_3(z) =\frac{1}{2}\left(
3 K_1(z) + K_2(z) - \gamma - \Xi + 2 \epsilon \right)
\end{equation}
with
\begin{equation}\label{evec3}
a_1 = \frac{-K_1(z) + K_2(z) + \gamma + \Xi}{4 \tau}=-a_2; \hspace{0.3cm}
%a_2 = -a_1; \hspace{0.3cm}
a_3 = a_4 = 1
\end{equation}
and
\begin{equation}\label{lamb4}
\lambda_4(z) = \frac{1}{2}\left(
3 K_1(z) + K_2(z) - \gamma + \Xi + 2 \epsilon \right)
\end{equation}
with
\begin{equation}\label{evec4}
a_1 = \frac{-K_1(z) + K_2(z) + \gamma - \Xi}{4 \tau}=-a_2; \hspace{0.3cm} 
% a_2 = - a_1; \hspace{0.3cm}
a_3 = a_4 = 1 \hspace{0.2cm}.
\end{equation}
In the right panels of Figs. \ref{fig2b} and \ref{fig3b} we plot the four lowest 
poles of the resolvent of Eq. (\ref{resolv}) with the effective 
Hamiltonian $H_{\rm eff}$ of Eq. (\ref{Heff}) as functions of $g$. 
In particular, in Fig. \ref{fig3b} curve 1 represents a solution of $z =\lambda_3(z)$,
curve 2 a solution $z = \lambda_2(z)$, curve 3 a solution of $z = \lambda_1(z)$,
and curve 4 a solution of $z = \lambda_4(z)$.
These four poles are compared with the four eigenvalues
%$\lambda_3$, $\lambda_2$, $\lambda_1$, $\lambda_4$ 
of the restricted model with at most one phonon
excitation of Sect. \ref{sect.restricted}, shown 
in the left panels of Figs. \ref{fig2b} and \ref{fig3b}. 

%Moreover, it is valid when the coefficients $K_1$ and $K_2$ are not singular. 
%For this to be true we also require $\omega_0 \gg U$ and $\omega_0 \gg g$.
%These conditions are valid for the deep and tight double-well potential. 

%\subsection{$J=0$}
In case there is no tunneling (i.e. $J=0$) we get degenerate states and only 
two different eigenvalues which cross each other are available (see Fig. \ref{fig2b}). 
The ground state thus has a cusp at $g=\sqrt{U\omega_0/2} \approx 7.1$ 
and coincides with the exact ground state of the Holstein-Hubbard
model for vanishing tunneling given in Eq. (\ref{ground0}). 
This was not the case, when we considered the exact solution
with only one phonon excitation in the previous section.

\begin{figure}[t]
\begin{center}
\includegraphics[width=7cm]{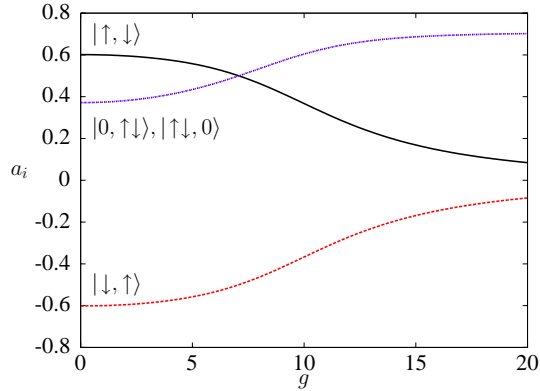}
\caption{\label{fig.vect}
Coefficients $a_i$ (see Eq. (\ref{expansion})) for the ground state of the 
effective Hamiltonian $H_{\rm eff}$ in Eq. (\ref{Heff}) (i.e. curve 1 in the right panel of 
Fig. \ref{fig3b}) with $J = 1$, $U = 2$ and $\omega_0 = 50$.
They cross very close to the crossing point of the eigenvalues in 
Fig. \ref{fig3b}. The domination of the singlet
states $|\!\uparrow,\downarrow\rangle$ at small couplings $g$ is changed by the 
domination of the dimer states for large values of $g$.}
\end{center}
\end{figure}

%\subsection{$J=1$}
A nonzero tunneling lifts the degeneracies and leads to a unique ground state. 
The eigenvalues for nonzero tunneling and for  $U=2$, $\omega_0=50$ are shown in 
Fig. \ref{fig3b}. 
The two lowest excited states  still  cross at around $g \approx 7.1$, 
while the ground state is unique and thus the system does 
not exhibit the transition discussed in Sect. \ref{sec2}. 
As a consequence of the polaron effect, the renormalized tunneling rates $\tau$, $K_1$, and $K_2$
vanish for large $g$. This implies for the eigenvalues the asymptotic behavior
\begin{equation}
\lambda_1\sim\lambda_4\sim-4g^2/\omega_0 , \ \ \ 
\lambda_2\sim\lambda_3\sim-2g^2/\omega_0
\ . 
\end{equation}
This is also visible on the right panel of Fig. \ref{fig3b}, 
while for the exact two sites problem with at most one phonon excitation 
(previous section), the eigenvalues do not merge for large coupling $g$ 
(cf. left panel of Fig. \ref{fig3b}).

We plot the coefficients $a_i$ (see Eq. (\ref{expansion})) for the ground state 
of the effective Hamiltonian in Fig. {\ref{fig.vect}}. 
There is a crossover from the domination by the singlet states 
$|\!\uparrow,\downarrow \rangle$ (at small coupling $g$ when the effective 
interaction is repulsive, $\gamma > 0$) to the domination of the doubly occupied states 
(at larger coupling $g$ when the effective interaction is attractive, 
$\gamma < 0$).

To discuss the consequences of the crossing on observable dynamical quantities 
we study in the next section the dynamics of the quantum states and the spin imbalance 
of the LFA. This investigation includes the spectral density of the model.

\section{Dynamics in a double-well potential}
\label{sect4}

%A double-well potential can be realized by superimposing two periodic potentials
%with different periodicity and the dynamics of one or two atoms can be easily
%studied. 

The description of the dynamics of our quantum system is based on the knowledge of
the energy levels, the initial quantum state and the overlap of the energy eigenstates
with the initial quantum state. In other words,
if the system with energy levels $E_j$ is prepared initially in state $|\Psi_0\rangle$, 
its quantum state $|\Psi_t\rangle$ evolves in time as  
\begin{equation}\label{tevol}
|\Psi_t\rangle=e^{-itH}|\Psi_0\rangle =\sum_j e^{-it E_j}
|E_j\rangle\langle E_j|\Psi_0\rangle
\ .
\end{equation}
The energy levels $E_j$ and the overlap with the initial state $\langle E_j|\Psi_0\rangle$
can be described by spectral density. This will be discussed in the next section.

\subsection{Spectral density}
\label{spectdens}

The return probability of the system to the initial state is calculated
from the inverse Laplace transform of the projected resolvent of Eq. (\ref{resolv})
as \cite{ziegler09} 
\begin{equation}\label{rp}
\langle \Psi_0|\Psi_t\rangle = 
\int_{\Gamma} e^{-i z t} \langle \Psi_0|(z-H_{\rm{eff}}(z))^{-1}_0|\Psi_0\rangle
\frac{dz}{2 \pi i} \hspace{0.2cm},
\end{equation}
where the contour $\Gamma$ includes all the poles of the resolvent.
The many-body spectral density is then given by
\begin{equation}\label{spectd}
\rho_\delta(E)=
- \frac{1}{\pi}{\rm Im} \langle \Psi_0 | \bigl(z - H_{\rm{eff}}(z)\bigr)_0^{-1}|\Psi_0\rangle
\hspace{0.2cm},
\end{equation}
where $z = E + i \delta$ and $\delta \ll 1$. For a finite dimensional Hilbert space
it is a rational function with poles $z = E_j \, (j=0, 1, 2, ...)$
\begin{equation}\label{spectda}
\rho_\delta(E)=
\delta\sum_j \frac{|\langle\Psi_0|E_j\rangle|^2}{(E-E_j)^2+\delta^2}
\hspace{0.2cm} .
\end{equation}
This expression represents Lorentzian peaks at positions $E_j$ whose heights
are $|\langle\Psi_0|E_j\rangle|^2/\delta$.
Plotting $\rho_\delta(E)$ as a function of $E$, we can clearly identify the poles $E_j$
of the resolvent $G_0$ (see Eq. (\ref{resolv})) and the overlap between the energy
state $|E_j \rangle$ and the initial state $|\Psi_0 \rangle$.
Here we calculate the spectral density for the initial state 
$|\Psi_0\rangle = |\!\uparrow,\downarrow\rangle$ for different coupling $g$. 
Then it should be noticed that the initial state is singly occupied and has no overlap
with the doubly occupied eigenstate of Eq. ({\ref{lamb1}). 
The results are shown in Fig. \ref{sdplots} for low energies.
We observe three peaks, which correspond to the energies shown in Fig. 
\ref{fig3b}. The central peak represents the dominant energy level for the dynamics. 
We notice the absence of one peak (corresponding to curve $3$
in Fig. \ref{fig3b}), since the state 
$\bigl(|\!\uparrow \downarrow,0\rangle + |0,\uparrow \downarrow\rangle\bigl)/\sqrt{2}$ 
can not be reached from the initial state $\Psi_0 = |\!\uparrow,\downarrow\rangle$.
We also observe that before ($g = 4$) and after ($g = 10$) the crossing of
the eigenvalues, there is one dominant central peak and two lower peaks. 
The characteristic frequencies of the dynamics %spin imbalance discussed previously 
are the differences of the energies of the central peak and the
other ones. Close to the crossing point ($g = 7.06$), the two other peaks are
symmetric with respect to the central one. Consequently, only one frequency appears
in the dynamics at the crossing point. This will also be seen in the spin imbalance of
the next section. For big values of $g$
we observe two large peaks which are very close and a small peak far from the central peak.
This implies a dominating small single frequency, as also found in the spin imbalance (cf. Fig. 
\ref{fig.g20}).
% of the spin imbalance be very small (difference between 
%the main peak and the closest one) and the other frequency be large 
%(see Fig. \ref{fig.g20}).

\begin{figure}[h!!]
\begin{center}
\begin{minipage}[b]{7cm}
\includegraphics[width=7cm]{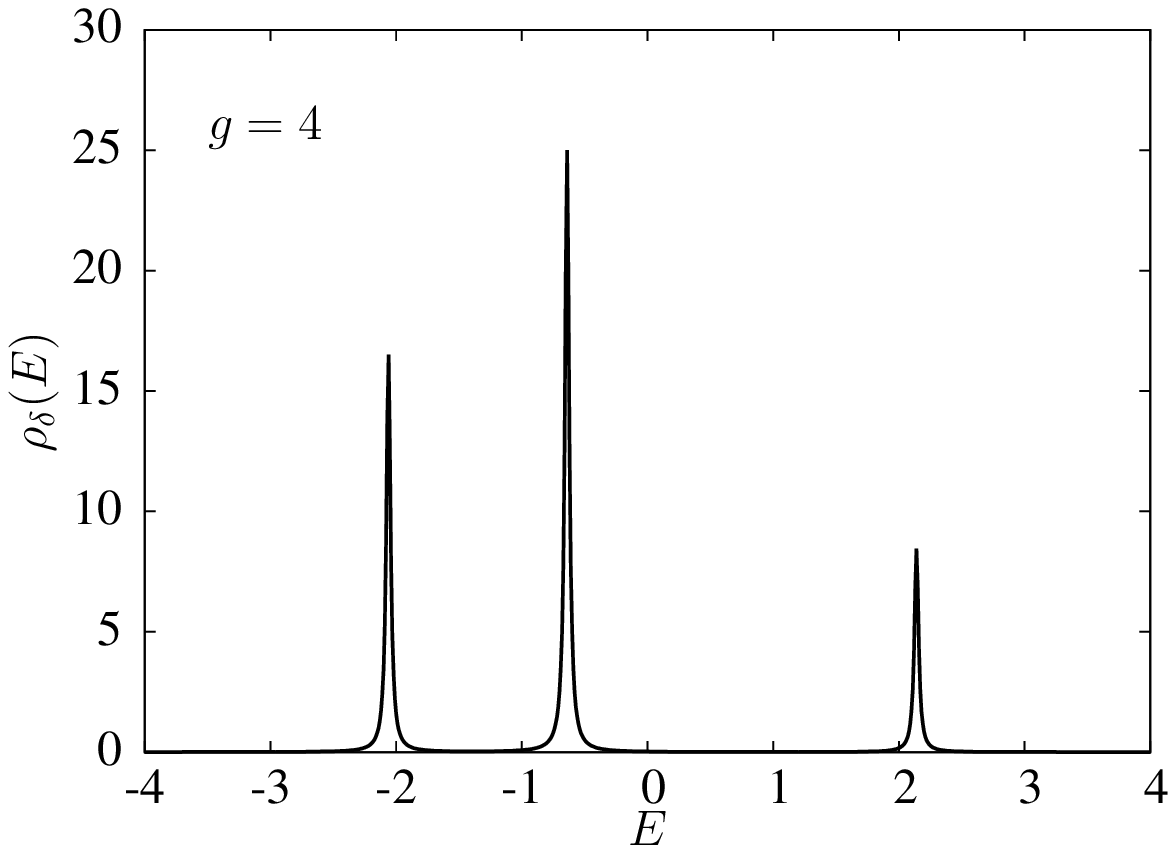}
%\includegraphics[width=\linewidth]{sdg4c.eps}
%\caption{g=4}
\end{minipage} 
%\hfill
\begin{minipage}[b]{7cm}
\includegraphics[width=7cm]{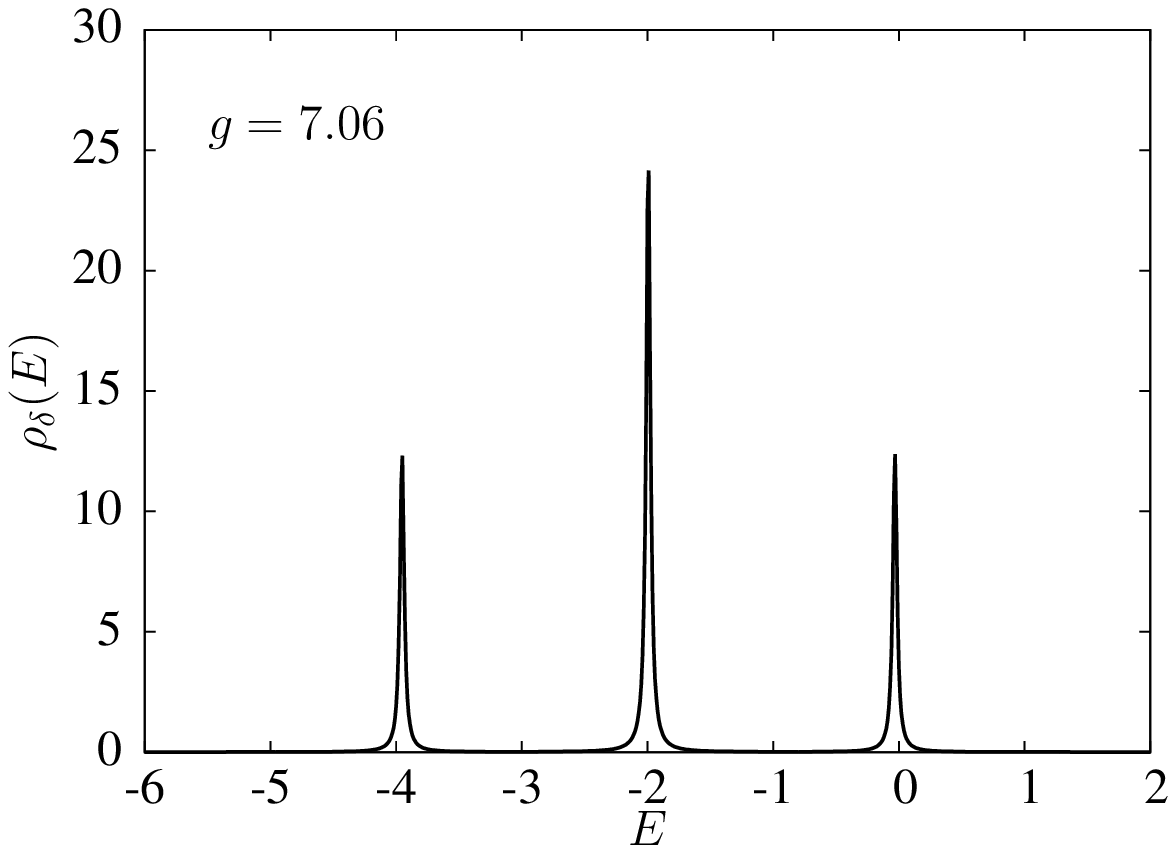}
%\includegraphics[width=\linewidth]{sdg7c.eps}
%\caption{g=7.06}
\end{minipage} 
\vfill
%\end{figure}
%\begin{figure}[!htb]
\begin{minipage}[b]{7cm}
\includegraphics[width=7cm]{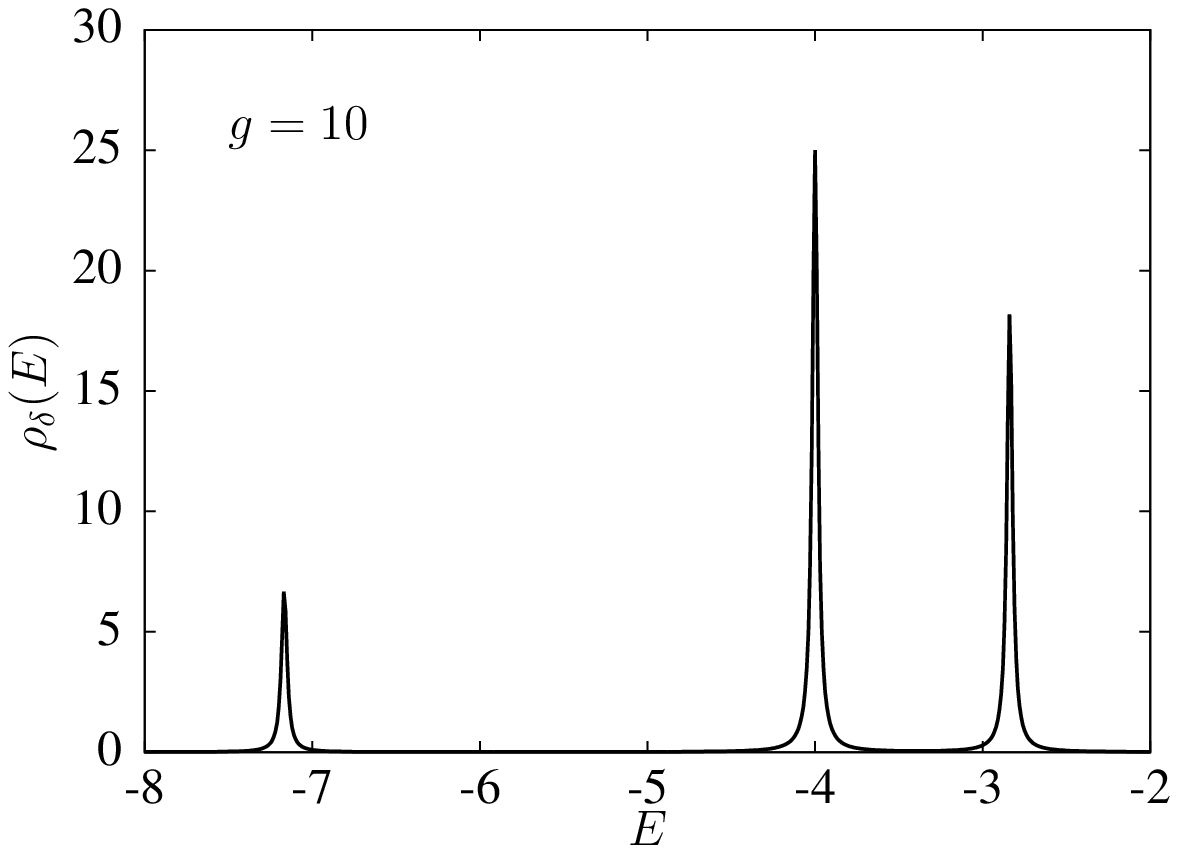}
%\includegraphics[width=\linewidth]{sdg10c.eps}
%\caption{g=10}
\end{minipage} 
%\hfill
\begin{minipage}[b]{7cm}
\includegraphics[width=7cm]{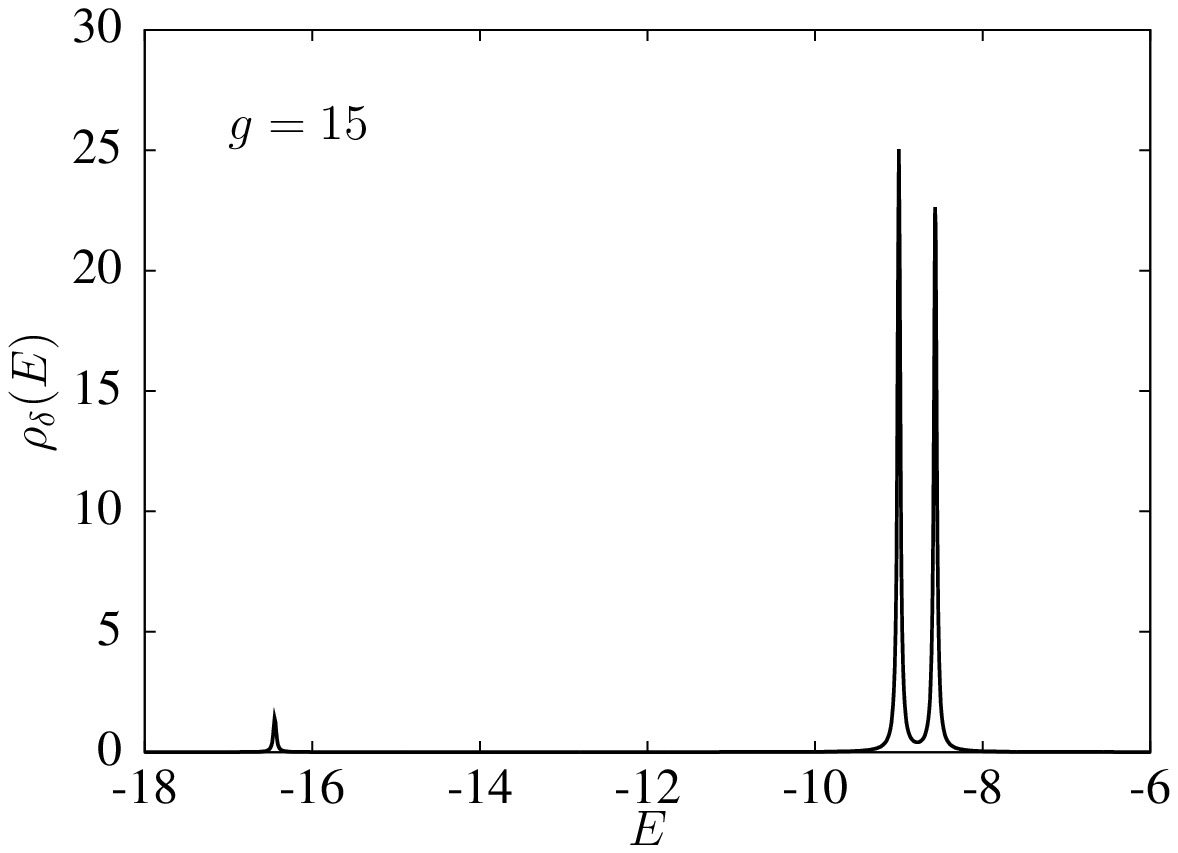}
%\includegraphics[width=\linewidth]{sdg15c.eps}
%\caption{g=15}
\end{minipage}
\caption{\label{sdplots} The spectral density in Eq. (\ref{spectda}) for the initial state
$|\Psi_0\rangle = |\!\uparrow,\downarrow\rangle$ and for $J=1$,
$U=2$, $\omega_0 = 50$, $\delta=0.02$ and different values of the coupling $g$. 
We observe three peaks because the fourth eigenstate is orthogonal to the initial state. 
The central peak represents the dominant energy level and for 
big values of $g$ the contribution of one peak is very small.
The frequencies of the spin imbalance are given by the difference between the dominant
energy level and the other ones.}
\end{center}
\end{figure}

\subsection{Spin imbalance}

A recent experimental study of the dynamics of two spin-1/2 atoms with strong 
repulsion in a double well has revealed that, using two singly occupied wells as
the initial state, the single occupation is static while the spin oscillates 
periodically between the two wells with two characteristic frequencies
\cite{bloch07,bloch08}. This observation has been interpreted by an effective
dynamics based on the Heisenberg model. The latter can be understood either within
a strong-coupling approximation of the underlying Hubbard model \cite{bloch08}
or within the recursive projection method for a general coupling \cite{ziegler09}. 
Experimentally this
has been seen by measuring the spin imbalance between the two wells
\begin{equation}\label{spin}
N_{1,2}(t)~=~\frac{1}{2} \langle \Psi_t |
n_{\uparrow 1} - n_{\downarrow 1} + n_{\downarrow 2} - n_{\uparrow 2} |
\Psi_t \rangle \hspace{0.2cm}.
\end{equation}
In our Holstein-Hubbard model we can vary the coupling $g$ between the LFA and the HA
to realize an additional interaction. We have already seen in the spectral density
that there is no overlap between the state of singly occupied wells and a state
of a doubly occupied well. From this point of view we expect a similar behavior
as found for the Hubbard model. However, there is the additional feature that we can
tune continuously the local atom-atom coupling $\gamma$ from an attractive to a repulsive
interaction. In this way we also reach a degeneracy point at which the effective 
interaction vanishes (i.e. $\gamma=0$).   
The existence of only three peaks in the spectral density of Fig. \ref{sdplots}
explains the fact that the spin imbalance is 
characterized by only two frequencies, i.e., the difference between the dominant 
energy level and the other levels.

%If the system is prepared initially in state $|\Psi_0\rangle$, it evolves 
%in time through the propagator $e^{-it
%\hat{H}}$: $|\Psi_t\rangle=e^{-it\hat{H}}|\Psi_0\rangle$. 
%Then the spin imbalance is defined as
%\begin{equation}\label{spin}
%N_{L,R}(t)~=~\frac{1}{2} \langle \Psi_t |
%n_{\uparrow L} - n_{\downarrow L} + n_{\downarrow R} - n_{\uparrow R} |
%\Psi_t \rangle.
%\end{equation}
For $g=0$ we get the Fermi-Hubbard model without phonon excitations which corresponds with
the above mentioned experiment. In this case,
if the initial state is $|\uparrow,\downarrow\rangle$, the dynamics of 
spin imbalance is characterized by two frequencies, 
$U/2 \left(\sqrt{(4J/U)^2+1}\pm 1\right)$ (see \cite{bloch08,ziegler09}).
The corresponding spin imbalance for $U=2$ is plotted in Fig. \ref{fig.FH}. 
\begin{figure}[t]
\begin{center}
\includegraphics[width=7cm]{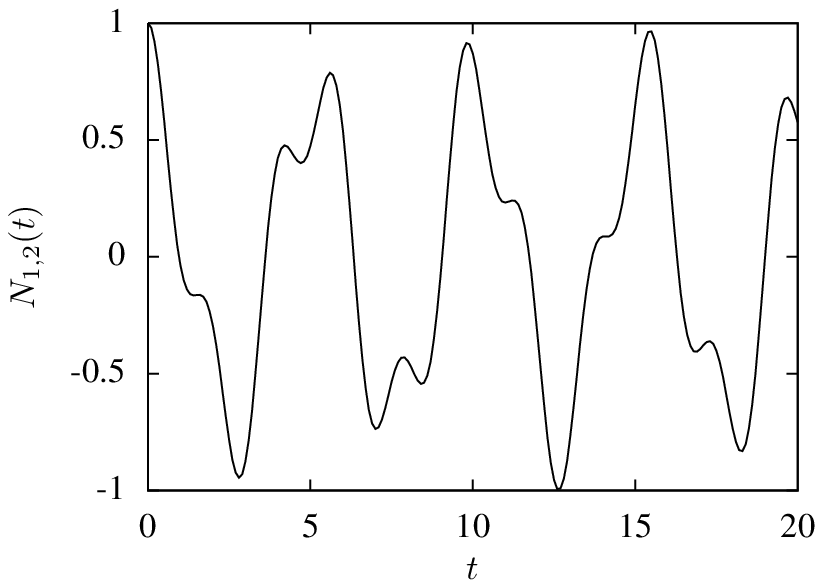}
\includegraphics[width=7cm]{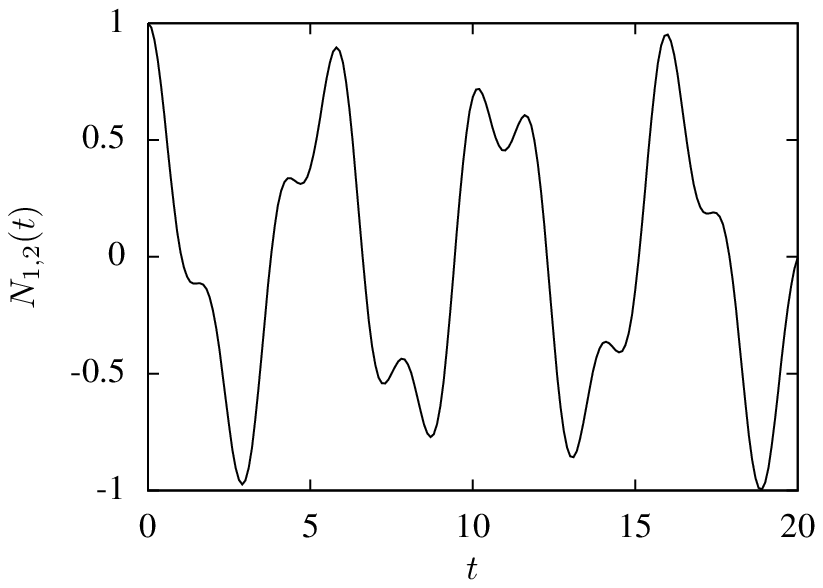}
%{g0.eps}
\caption{\label{fig.FH}
Spin imbalance $N_{1,2}(t)$ without inelastic scattering ($g=0$, left panel)
and with strong inelastic scattering ($g=10$, right panel) for $J = 1$,
$U = 2$ and $\omega_0 = 50$. There are two modes with frequencies 
$\omega_1 \approx 3.24$ and $\omega_2 \approx 1.24$ (left panel) and
$\omega_1 \approx 3.17$ and $\omega_2 \approx 1.16$ (right panel)
contribute to the dynamics.}
\end{center}
\end{figure}
\begin{figure}[t]
\begin{center}
\includegraphics[width=7cm]{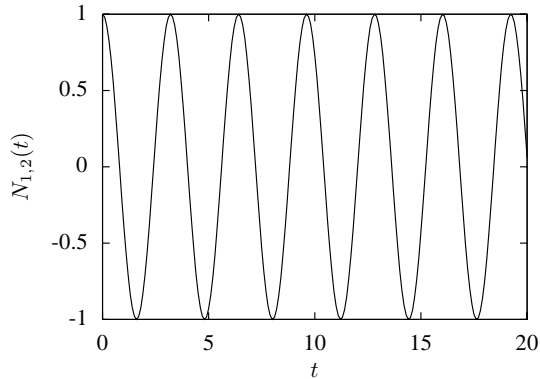}
%{gc.eps}
\caption{\label{fig.cross}
Spin imbalance $N_{1,2}(t)$ at the level crossing $g=7.06541$
for $J = 1$, $U = 2$ and $\omega_0 = 50$.
In this case only one mode with frequency $\omega_1 \approx 1.96$ 
contributes to the dynamics.}
\end{center}
\end{figure}

For nonzero coupling $g$ the dynamics is affected by the presence of 
phonon excitations but it is still characterized by two frequencies, 
which are the differences between the second and the first
and the second  and the fourth curves in Fig. \ref{fig3b}, respectively, 
i.e., $\omega_1 = \lambda_2 - \lambda_3$ and 
$\omega_2 = \lambda_2 - \lambda_4$.
Consequently, at the crossing
%As the result at the crossing 
the dynamics of spin imbalance shows only one frequency 
since the differences between the curves in this case 
are equal, as depicted in Fig. \ref{fig.cross}. 
For larger couplings $g$ two frequencies component appear again as it is shown 
in the right panel of Fig. \ref{fig.FH}. 
Thus measuring the difference between the two frequencies, 
$\Delta = |\omega_2 - \omega_1|$   %, which contribute to the spin imbalance dynamics 
provides a method to detect the crossing point experimentally: $\Delta$ vanishes 
at the crossing point as shown in Fig. \ref{fig.w1-w2} and the spin imbalance 
is characterized by only one frequency. 
Increasing the coupling $g$ further leaves a low frequency component with almost 
full amplitude and additional high-frequency modulation with small amplitude. 
This is depicted in Fig. \ref{fig.g20}.
The low frequency component oscillates with the frequency 
$\omega_2 \approx 4 \tau^2/|\gamma|$,
where $\tau$ is given in Eq. (\ref{ren1}) and $\gamma = U - 2 g^2/\omega_0$.
Thus $\omega_2 \rightarrow 0$ as $g \rightarrow \infty$,
which is a direct consequence of the polaron effect.
%%%%%%%%%%%%%%%%%%%%%%%%%%%%%%%%%%%%%%%%%%%%%%%%%%%%%%%%%%%%%%%%%%%%%%%%%%%%%%%%

%%%%%%%%%%%%%%%%%%%%%%%%%%%%%%%%%%%%%%%%%%%%%%%%%%%%%%%%%%%%%%%%%%%%%%%%%%%%%%%%
\section{discussion and conclusion}

%-- comparing 1 phonon vs. eff. Hamiltonian: a) J=0 Figs. 2,4 b) J=1 Figs. 3,5. 

At low energies,
the restricted model with at most one phonon excitation per well has qualitatively
the same behavior as the model with many phonon excitations, described by the effective
Hamiltonian in Eq. (\ref{Heff}). This is presented in Figs. \ref{fig2b} and \ref{fig3b}, where the four lowest levels
are plotted for both cases. The main difference, however, is that the Hilbert space
of the model with many phonon excitations is much larger. Consequently, there are many 
excited states with energies higher than those shown in Figs. \ref{fig2b}, \ref{fig3b}.
However, these states are not considered here because of their high energies.
Due to the matrix elements $K_1$ and $K_2$ of the effective Hamiltonian in Eq. (\ref{Hmatrix}),
these higher levels are closely related to harmonic oscillator levels with frequency
$\omega_0$.

Without tunneling (i.e. $J=0$) there is a change of the ground state from single 
occupancy of the wells (weak coupling $g$) to double occupancy of one well (strong coupling $g$). 
This reflects the sign change of the effective coupling $\gamma=U-2g^2/\omega_0$. In the presence
of tunneling (i.e. $J\ne 0$) the ground state, given by the coefficients of Eq. (\ref{evec3}),
changes smoothly upon a change of the coupling $g$. Its energy is the lowest solution of $\lambda_3(z)=z$, 
where $\lambda_3(z)$ is defined in Eq. (\ref{lamb3}). There is a transition due to the crossing of 
the first and second excited level though (cf. Fig. \ref{fig3b}).

%\section{Conclusions}

In conclusion, we have studied an atomic mixture of two heavy atoms and two light spin-1/2 
fermionic atoms in a double-well potential, where the heavy atoms are subject to 
local harmonic oscillator potentials of the wells.
This is modeled using the Holstein-Hubbard Hamiltonian,
which is  the simplest system that mimics the presence of phonons in a solid.
We have applied the recursive projection method, which
reduces the complexity of the full Hilbert space and leads to 
an effective fermionic Hamiltonian. 
We have found a transition for the the light fermions from singly occupied wells to 
doubly occupied wells as the coupling between heavy and light species is increased.
This transition is manifested by the crossing of the second and third eigenvalue
of the effective Hamiltonian. 
Moreover, the coupling between the light and the heavy atoms renormalizes the tunneling 
of light fermions between wells, 
which reflects the polaron effect. The dynamics is dominated by a spectral density with three
peaks. This implies for the spin imbalance dynamics of the light atoms a periodic behavior with
two characteristic frequencies. These frequencies coincide at the crossover of the two lowest
excited states. Thus the oscillating behavior of the spin imbalance can be used to detect the
crossing point experimentally.

\begin{acknowledgments}
This work was supported by Coordena\c c\~ao de Aperfei\c coamento de Pessoal 
de N\'ivel Superior (CAPES) and by the Deutscher Akademischer Austausch Dienst (DAAD).
\end{acknowledgments}

\begin{figure}[t]
\begin{center}
\includegraphics[width=7cm]{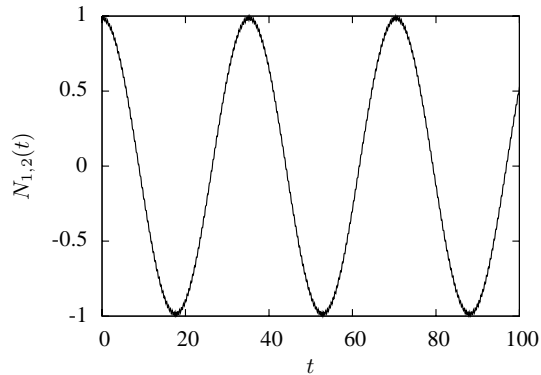}
\caption{\label{fig.g20}
Spin imbalance $N_{1,2}(t)$ for $g=20$, $J = 1$,
$U = 2$ and $\omega_0 = 50$. Large coupling $g$ leaves a slow component with almost 
full amplitude
and an additional high-frequency modulation with small amplitude. The slow component
oscillates with the frequency $\omega_2 \approx 4 \tau^2/|\gamma|$. 
This is the direct consequence of the polaron effect, since
it causes the single fermion tunneling decay exponentially, 
$\tau=Je^{-g^2/\omega_0^2}$.}
\end{center}
\end{figure}
\begin{figure}[h!!]
\begin{center}
\includegraphics[width=7cm]{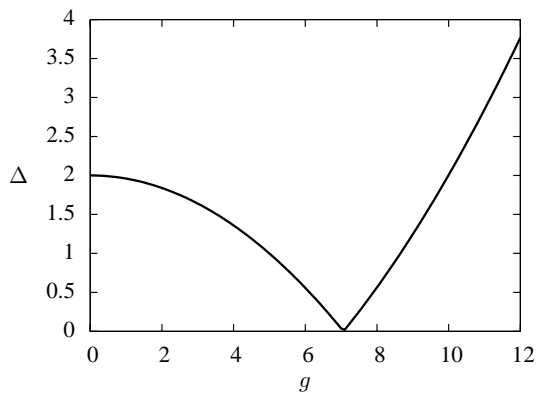}
%{diff2.eps}
\caption{\label{fig.w1-w2} Difference
$\Delta = \bigl| \omega_2 - \omega_1 \bigr|$
between the frequencies of the modes contributing to the dynamics of spin imbalance. 
The difference disappears at the crossing point.}
\end{center}
%\label{fig.w1w2}
\end{figure}
%%%%%%%%%%%%%%%%%%%%%%%%%%%%%%%%%%%%%%%%%%%%%%%%%%%%%%%%%%%%%%%%%%%%%%%%%%%%%%%%

%\newpage

\bibliographystyle{ieeetr}
\bibliography{ref}

\end{document}